\documentclass[a4paper,10pt]{article}
\usepackage[margin=1in]{geometry}
\usepackage{authblk}
\usepackage{booktabs}
\usepackage{amsmath}
\usepackage{url}
\usepackage{graphicx}
\usepackage{ntheorem}
\theoremseparator{:}

\usepackage[english]{isodate}
\newcommand{\ra}[1]{\renewcommand{\arraystretch}{#1}}

\title{A network approach to cartel detection in public auction markets}

\author[1,2]{Johannes Wachs}
\author[2]{J\'anos Kert\'esz}

\affil[1]{Chair for Computational Social Sciences and Humanities, RWTH Aachen}
\affil[2]{Department of Network and Data Science, Central European University}

\begin{document}

\flushbottom
\maketitle

\begin{abstract}

Competing firms can increase profits by setting prices collectively, imposing significant costs on consumers. Such groups of firms are known as cartels and because this behavior is illegal, their operations are secretive and difficult to detect. Cartels feel a significant internal obstacle: members feel short-run incentives to cheat. Here we present a network-based framework to detect potential cartels in bidding markets based on the idea that the chance a group of firms can overcome this obstacle and sustain cooperation depends on the patterns of its interactions. We create a network of firms based on their co-bidding behavior, detect interacting groups, and measure their cohesion and exclusivity, two group-level features of their collective behavior. Applied to a market for school milk, our method detects a known cartel and calculates that it has high cohesion and exclusivity. In a comprehensive set of nearly 150,000 public contracts awarded by the Republic of Georgia from 2011 to 2016, detected groups with high cohesion and exclusivity are significantly more likely to display traditional markers of cartel behavior. We replicate this relationship between group topology and the emergence of cooperation in a simulation model. Our method presents a scalable, unsupervised method to find groups of firms in bidding markets ideally positioned to form lasting cartels.

\end{abstract}


\section*{Introduction}
Cartels have been studied by economists since Adam Smith, who wrote that ``people of the same trade seldom meet together, even for merriment and diversion, but the conversation ends in a conspiracy against the public, or in some contrivance to raise prices.''~\cite{smith2003wealth} Competition between firms decreases profits, and so they have incentive to collude by setting prices or production collectively~\cite{marshall2012economics}. Three significant forces make running such a cartel difficult. First, they are generally illegal and are aggressively policed by competition authorities, presenting a significant obstacle to coordination. The second challenge is that cartels face a classic collective action problem: individual cartel members have short-term incentive to deviate from the collusive agreement. In the language of game theory, defection from the cartel is a dominant strategy when a single round is played. Finally, effective collusion requires unanimity, meaning that a single defector can significantly diminish the profits of the other cooperating firms.

Even though cartels face these internal and external threats to their stability there are many examples of cartels in a variety of industries from financial products~\cite{abrantes2012lessons} to vitamins~\cite{harrington2011private} operating successfully for long periods of time. Past research estimates that on average cartels increase prices by 20-30\%~\cite{connor2006cartel} and that in a given year a functioning cartel has a roughly 10\% chance of being discovered~\cite{combe2008cartels}, indicating that cartels are costly and that deterrence can be improved. In this paper we propose a novel approach to the issue of detecting cartels in the specific context of public auction markets using network methods, built on the idea that the network of firm-firm interactions can reveal hot-spots in which cooperation is easier to sustain. Groups of firms in these distinguished positions interact repeatedly and exclusively among themselves, creating ideal conditions to overcome the internal collective action problem and to engage in bid-rigging, a common form of anti-competitive behavior in these markets.

Economists have long studied the market conditions under which cartels thrive, namely how they enable a cartel to overcome the problems of ``coordination, cheating, and entry''~\cite{levenstein2006determines}. For example coordination is easier in a smaller group with homogeneous firm sizes~\cite{compte2002capacity}, while frequency of interaction facilitates punishment of defectors, and high costs to entry insulate the cartel from outsiders~\cite{motta2004competition}. Another perspective on collusion is to consider that firms are playing a repeated prisoner's dilemma (PD) game~\cite{kofman1996aprisoner}. When choosing to compete (equivalently defecting from the cartel in the language of the PD) firms charge low prices, while when they collude (equivalently cooperate), they charge high prices. Collective profit is maximized if everyone colludes, but when colluding firms are undersold by even a single competing firm they fare badly~\cite{rapoport1974prisoner}. When players play multiple rounds of the PD, Axelrod demonstrated that cooperation can emerge as a winning strategy through learning and imitation~\cite{axelrod1981evolution,dixon2002axelrod}. In the PD and many other games in which collectively optimal actions are personally costly to players, altruistic cooperation emerges under a variety of conditions through mechanisms such as reciprocity and the altruistic punishment of defectors or cheaters~\cite{bowles2011cooperative}.

Just as certain market conditions are known to favor cartels, researchers have observed that when players of the PD are arranged in some space or network which restricts their potential interactions, the potential for the emergence stable cooperation crucially depends on the structure of the space~\cite{nowak1993spatial,santos2008social,perc2013evolutionary,battiston2017determinants}. For example, correlations in the spatial distribution of agents have been shown to facilitate cooperation~\cite{szamado2008effect}. To the best of our knowledge, this observation that local correlation of interactions have significant influence on the emergence of cooperation has not been applied to the problem of detecting cartels. This paper proposes focusing the search for cartels on groups occupying ideal positions in the competitive landscape for the emergence of cooperation.

We propose to apply network science methods to identify groups of intensely interacting firms in a market and to screen them for collusive potential, based on their network topology and how it may facilitate collusion. We focus on the specific case of collusion in public contracting markets, in which public bodies buy goods and services from private firms. These markets are vulnerable to collusion because of the inelasticity and regularity of government's demand for certain goods~\cite{pesendorfer2000,huschelrath2014cartel}. They are also large, accounting for between 10-20\% of GDP in the OECD~\cite{oecdprocurement}. Contracts are commonly awarded via auction to the lowest bidder. In these markets cartels often engage in bid-rigging, coordinating their bids to mask their agreement to avoid competition~\cite{porter1993detection}. The US Department of Justice highlights bid-rigging as one of the primary modes of anti-competitive behavior in public markets, besides collective price-setting and market-allocation~\cite{doj2005price}.

Specifically, we use data on firms bidding for contracts to map in public contracting markets as networks of competing firms. We argue that such a network represents an embedding of the firms into a space which describes the competitive landscape of their industry or location, including its geography, technology, and scale. Within such co-bidding networks, we detect groups of firms whose local network topology are naturally conducive to sustaining collusion. Our findings suggest that certain topological features may be necessary for cartels to successfully operate in the long-run.

Previous work on cartels has considered co-bidding networks of firms~\cite{toth2014toolkit,reeves2017bid,morselli2018network}, but using network topology to detect groups of firms within markets is a relatively new idea. For example Conley and Decarolis~\cite{conley2016} use an agglomerative clustering method to group firms based on their bidding behavior. In recent work on cartel screening, Imhof et al. have used patterns of bidding interactions between firms to study cartels~\cite{imhof2018screening}, though do not consider interactions between cartel and non-cartel firms. In general data on bids in public procurement are not made public, presenting a major obstacle to research on bid-rigging. This information is kept secret because it can be useful to firms engaged in collusion~\cite{albanopreventing}. In fact the OECD recommends that issuers of public procurement contracts do not publish information on losing bidders and bids for this reason~\cite{oecdbid}. The responsible authorities certainly have access to such information.

More broadly, network methods have been fruitfully applied to problems in criminology including corruption~\cite{fazekas2017corruption,ribeiro2018dynamical,wachs2019social}, the mafia~\cite{agreste2016network}, and the evolution of criminal behavior in society~\cite{rostami2015complexity}.

We first map the co-bidding market of the suppliers of public school milk in 1980s Ohio containing a known cartel case~\cite{porter1999ohio}. We note that the cartel firms occupy a distinguished position in the network: they frequently interact with one another, forming coherent links, and are relatively isolated from outside firms, forming an exclusive group. We then turn to a dataset of bids on nearly 150,000 contracts awarded in the Republic of Georgia from 2011 to 2016 worth roughly 5 billion US dollars. Using a greedy, bottom-up algorithm to detect overlapping groups of interacting nodes, we find that groups with cohesive and exclusive interactions have higher prices, are less likely to sue each other, and are more likely to have low variance in their bids and prices - classic screens for cartel behavior used by competition authorities around the world~\cite{abrantes2006variance}. Finally, we simulate a market in which firms compete for randomly placed contracts when they are close in proximity, introducing spatial correlations to interactions. Firms see their competitors for the contract, and decide to compete or collude based on the previous actions of their partners and the frequency with which they meet them. In the resulting co-bidding network, detected groups with coherent and exclusive links successfully collude with much higher frequency.

\section*{Results}
Our framework to find groups of firms that may be engaging in collusion consists of several steps. First we extract the co-bidding network of firms in a market, connecting two firms by an edge with a weight that increases as they more frequently bid for the same contract. We then identify groups of firms which frequently bid for the same contracts using a modified version of a popular overlapping community detection algorithm~\cite{lancichinetti2009detecting}. The method is greedy, and the function to merge nodes into groups has a penalty term for the number of nodes included, insuring that the groups detected remain small relative to the size of the market. Finally, we calculate topological features of the groups: their \textit{coherence} and \textit{exclusivity}. We suggest sustained collusion is more likely to emerge among high coherence and exclusivity groups because they offer the ideal conditions for firms to learn to cooperate and trust one another. We find evidence of this phenomenon in three settings: a dataset of school milk contracts with a known cartel, a dataset of virtually all contracts awarded in the Republic of Georgia over several years, and in a simulation model of contracting markets with spatial correlations. 

\subsection*{The 1980s Ohio School Milk Market}
We first analyze bidding data from the market for public school milk in 1980s Ohio~\cite{porter1999ohio}. Every summer school districts called for bids from dairies to provide school milk for the following academic year. Firms submitted sealed bids quoting a price in cents per pint. In 1993 representatives from two firms confessed to colluding with a third firm to rig bids for contracts in the Cincinnati area as part of a settlement. The third firm eventually settled out of court, paying significant civil penalties.

Previous work by Porter and Zona highlights irregularities in the bidding behavior of the suspected cartel firms compared to the rest of the market~\cite{porter1999ohio}. Exploiting specific features about the market for school milk, the authors created an econometric model to predict the bids of firms on contracts, including information on the capacity of firms, the specifications of the bids (i.e. whether drinking straws were required), and the physical distance between the firm and school. They found that the bids submitted by cartel members were often decreasing in distance - a highly suspicious fact given that a major cost in the supply of school milk is its transportation.

\begin{figure}[t]
\centering
\includegraphics[width=\textwidth]{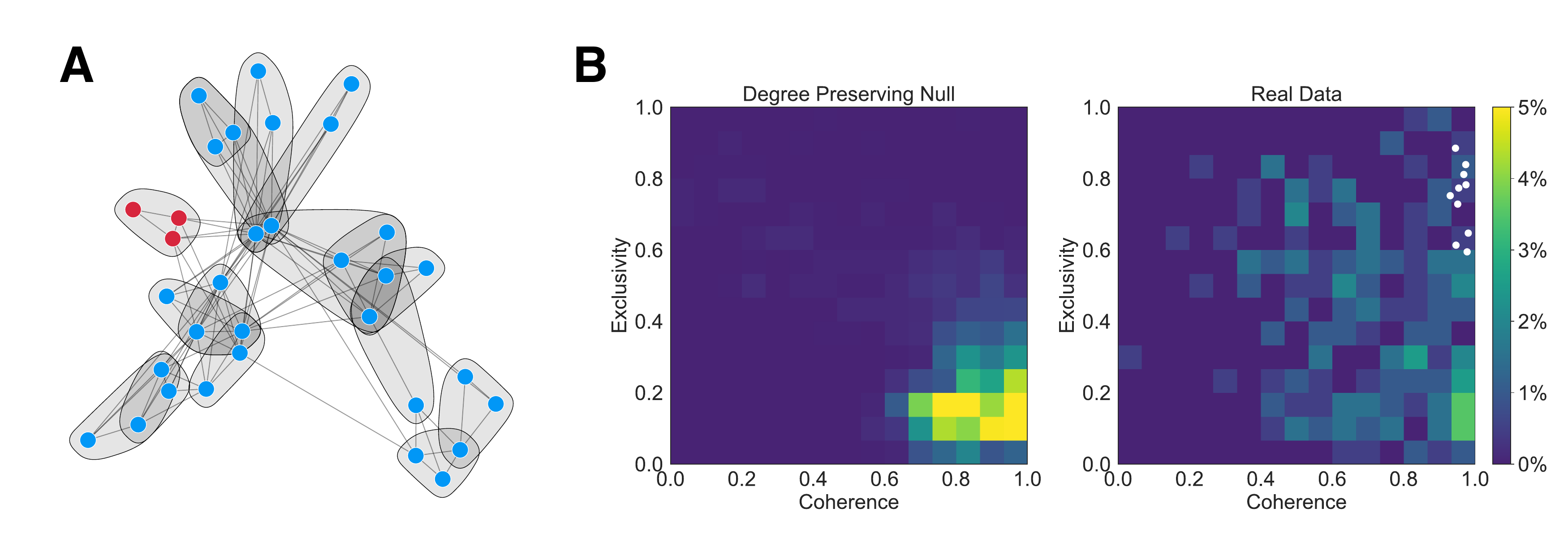}
\caption{A. Ohio school milk market co-bidding network, 1986. Overlapping groups are detected using our algorithm. Red nodes are member of the alleged cartel operating near Cincinnati. We exclude firms participating in less than 3 auctions for the purposes of visualization. B. Coarsened two-dimensional histograms of groups detected in all Ohio school milk networks 1980-1990 in the coherence-exclusivity space. The first plot shows the distribution of the groups detected in 100 bid-degree preserving null models of each of the 10 years. The second plot shows the real distribution of groups. The cartel group's position is marked by white circles. The cartel group has both high exclusivity and coherence.}
\label{fig:ohio_overlap}
\end{figure}

For each year from 1981-1990, inclusive, we created the co-bidding network of firms, connecting two firms based on the similarity of their bidding behavior. We apply our method to detect overlapping groups of interacting firms. We use a force layout algorithm to visualize the network in subplot A of Figure~\ref{fig:ohio_overlap}, highlighting the cartel firms in red and outlining the detected groups.  For each group we calculate its coherence, the ratio of the geometric to arithmetic means of its edge weights~\cite{onnela2005intensity} and its exclusivity, the ratio of strength within the group to the total strength of nodes in the group (including edges leaving the group). As features of groups of firms, coherence captures the consistency and intensity of interactions among firms in the group, while exclusivity quantifies the extent to which group interactions happen in isolation from the rest of the firms in the broader market.

We plot the distribution of groups across all ten years in the coherence-exclusivity space in subplot B of Figure~\ref{fig:ohio_overlap}. In the first plot we show the distribution groups detected in 100 null models for each year in which bidding behavior was randomized. Specifically, the null model shuffles bidders between contracts, such that each firm bids on the same number of contracts and each contract receives the same number of bids. In the second plot we show the observed distributions, indicating the position of the cartel firms (which our group detection algorithm identified as a group in each year) with white circles. We note two phenomena: the first is that groups in the empirical network have significantly higher exclusivity and coherence than what would be expected if the bids were random, while the second is that the high coherence and exclusivity regime is sparsely populated in both the empirical data and the null model.

\subsection*{Georgian Public Procurement Markets}
We now turn to data from a much larger procurement market covering a wide range of goods and services. Specifically we collected virtually all public contracts from the Republic of Georgia from 2011 to 2016. The data consists of nearly 150,000 contracts bid on by nearly 15,000 unique firms with total value roughly five billion US dollars. As with the Ohio dataset, we observe the bids and bidder identities for each contract. Rather than cluster our data by contract-level product type, we proceed by analyzing the whole network, arguing that firms participate in many markets and that any categorization of firm into market on the basis of firm or contract metadata would exclude many interactions between firms in adjacent markets. We visualize the co-bidding network for one year of data in the supplementary information, observing that most contracts are awarded to firms in a densely connected giant component of competitive activity.

We apply our method to detect overlapping groups in the whole market and calculate their coherence and exclusivity for each year. In the analysis that follows we consider only groups of firms identified from the co-bidding network that exclusively bid on at least 30 contracts in a given year in order to focus on significantly interacting firms. Our findings are robust to a range of cutoffs, which we report in the SI.

To compare our data against a plausible null model, we created randomized networks from data by shuffling the contracts firms bid on within specific product classes. This insures that firms bidding exclusively on school milk contracts do not bid on software consulting contracts in our null model. We use the resulting distributions of group cohesion and exclusivity from the null model to create thresholds for labeling groups from the empirical network as suspicious. We consider a group from the empirical co-bidding network suspicious if its coherence and exclusivity exceed the 80th percentile of coherence and exclusivity of groups in the null model in the same year. In the supplementary information we apply the same threshold to classify Ohio groups as suspicious and find that it consistently detects the cartel group with a low false positive rate.

\begin{figure}[t]
\centering
\includegraphics[width=\textwidth]{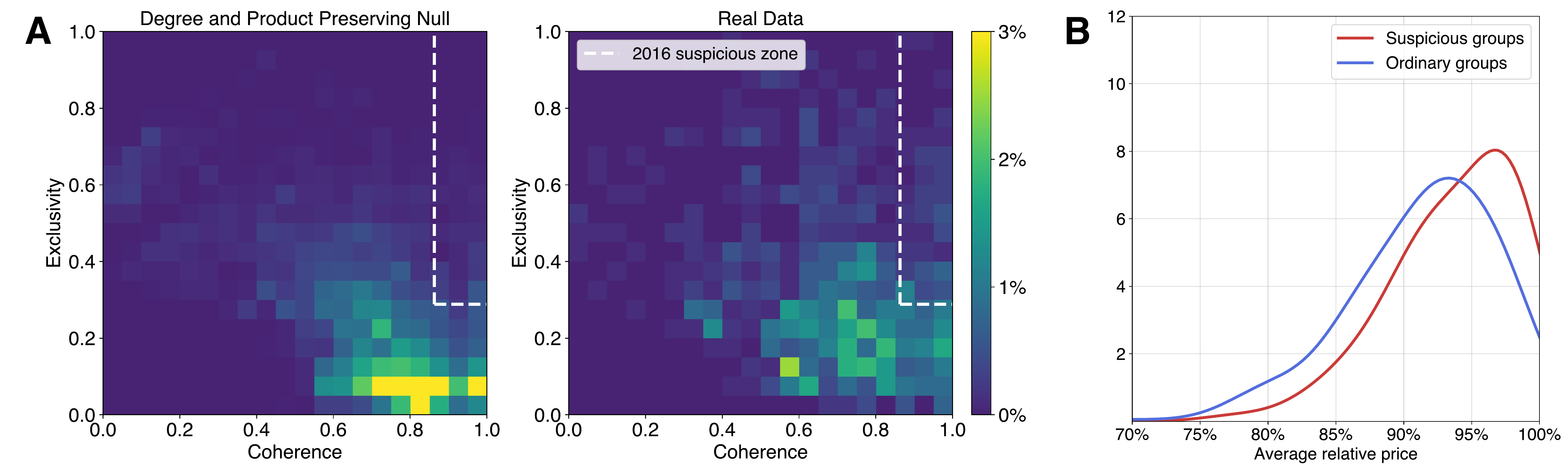}
\caption{A. The distribution of groups in the cohesion-exclusivity feature space detected in a product-type and bid-degree preserving null model compared with groups detected in observed data from Georgian procurement markets from 2011-2016. We label groups of firms as suspicious if its coherence and exclusivity are in the 80th percentile of the null model outcomes of the year in which they are detected. We highlight 2016's suspicious zone in white.  B. Distribution of average relative prices of contracts bid on by suspicious groups and ordinary groups, 2011-2016. Suspicious groups consistently win more expensive contracts.}
\label{fig:ga_results}
\end{figure}

We visualize the distributions of groups in the coherence and exclusivity space for the randomized and empirical data in subplot A of Figure~\ref{fig:ga_results}. We plot the data from all years in the same visualization, and highlight the suspicious zone of high coherence and exclusivity calculated for 2016.

As there are no confirmed ground truth cartels in our dataset (the Competition Authority of Georgia confirmed to us via direct correspondence that there have been no confirmed cartel cases in public procurement markets in Georgia~\cite{priv_comm_GCA}), we validate our claim that groups in the suspicious zone are operating under conditions that facilitate collusion using four measures. First we consider the average cost of contracts won by the group when they were the only participants. As contracts are announced with a reserve price, we can scale each contract's cost outcome to enable comparisons between contracts. We plot the distribution of relative prices for contracts won by suspicious groups versus all other groups in subplot B of Figure~\ref{fig:ga_results}. We confirm that groups in the suspicious zone are winning more expensive contracts, confirmed by a Mann-Whitney U test shown in Table~\ref{tab:ga_stats_table}.

Next, we calculated two price and bid based screens for collusion from the literature. The first is the price coefficient of variation of contracts won by the group~\cite{abrantes2006variance}, measuring the extent a group's prices are both high and stable. This screen is based on the theoretical observation that when prices are set collectively, it is costly to coordinate price changes~\cite{lacasse1995bid}. It aligns with empirical observations of real cartels~\cite{bolotova2008impact} and has been used extensively by competition authorities~\cite{abrantes2012lessons}. Specifically, the price coefficient of variation $CV_{price}^{G}$ of a group $G$ is defined in terms of the average cost of contracts $C$ cornered by the group, $\mu_{C}^{G}$, and the standard deviation  $\sigma_{C}^{G}$:

$$CV_{price}^{G} =\dfrac{\sigma_{C}^{G}}{\mu_{C}^{G}} $$

The second cartel screen we apply is the average of the coefficient of variation of bids on each contract for which only group firms submitted bids~\cite{imhof2017simple}. Previous research has shown that the fake bids submitted by losing members of the cartel tend to closely hug the winning bid. For each contract $c$ bid on exclusively by members of a group, we calculate the coefficient of variation in the bids:

$$CV_{bidding}^{c} =\dfrac{\sigma_{c}}{\mu_{c}} $$
We average over all contracts $C$ cornered by a group $G$ to obtain its bidding coefficient of variation:

$$CV_{bidding}^{G} =\mu_{c \in C}\left(CV_{bidding}^{c}\right) $$
We say that a group of firms has a low bidding coefficient of variation if it is less than one standard deviation below the market average. A Mann-Whitney U test, shown in Table~\ref{tab:ga_stats_table}, indicates that groups in the suspicious zone are significantly more likely to have lower $CV_{bidding}^{c}$ and $CV_{price}^{G}$. 

We carry out one more test of our method using data on bid protests. Bid protests are legal actions by firms against contracts awarded by procurement authorities. Firms can protest, for example, if the contract was not advertised in the proper venue, or if they believe criteria to participate in an auction unfairly excluded them. We collected data on which firms protested which contracts, including the firm to which the contract was awarded. We argue that colluding firms would never protest the contracts won by their cartel partners, while one may expecting intensely competing firms to frequently protest each others' winnings. For each group we check if any contract awarded to a group member was protested by another group member that year. We find that suspicious groups are half as likely to have such internal protests - a statistically significant difference shared in Table~\ref{tab:ga_stats_table}. 

\begin{table}[]
\ra{1}
\begin{tabular}{@{}lrrcrrclr@{}}
& \multicolumn{2}{c}{Suspicious Groups} & \phantom{abc}& \multicolumn{2}{c}{Ordinary Groups} & \phantom{abc} & \multicolumn{2}{c}{Differences} \\
\cmidrule{2-3} \cmidrule{5-6}  \cmidrule{8-9} 
&\textit{Mean} & \textit{Std. Dev.} &&\textit{Mean} & \textit{Std. Dev.} &&\textit{M-W U} & \textit{p-value}\\ \midrule
Avg. Relative Price & 0.938   & 0.046 & & 0.914 & 0.053 & & 30211\textsuperscript{***} & $<$ 0.001  \\
Avg. $CV_{price}^{G}$& 0.098  & 0.055 & & 0.117 & 0.059 & & 33470\textsuperscript{***} & $<$0.001 \\
Avg. $CV_{bidding}^{G}$ & 0.047   & 0.056 & & 0.055 & 0.038 & & 32306\textsuperscript{***} & $<$0.001 \\
In-group Bid Protest Rate& 0.134   & 0.341 & & 0.237 & 0.425 & & 37516\textsuperscript{*} & 0.011 \\
\bottomrule
\end{tabular}
\caption{Cartel screens applied to suspicious and ordinary groups of firms detected in the Georgia procurement market, 2011-2016. Cartel groups have higher average relative prices, are more likely to have a low average coefficient of variation on bids for a contract, and are less likely to legally protest the winnings of other group members. Mann-Whitney U: * $p < .05$, ** $p <.01$,*** $p <.001$ }
\label{tab:ga_stats_table}
\end{table}

Suspicious groups detected by our methods are more likely to manifest the four collusive markers we have measured than their non-suspicious counterparts. Though this is no proof of collusion, it does indicate that many of the groups of firms that competition authorities might be interested in investigating based on their behavioral patterns exist in the same high coherence and exclusivity zone as the Ohio school milk cartel. In the next section we present a simple simulation model of a procurement market with spatial correlations which replicates our observation that collusion is more common among cohesive and exclusive groups.

\subsection*{Simulation Model}
We simulated a market of interacting firms placed uniformly at random in the unit square. The location of firms can be interpreted as their physical location or as a more abstract position in a space of product similarities (for instance firms supplying computer hardware might be closer to one another). Contracts, also located randomly, attract bids from nearby firms, introducing spatial correlation to the interactions between firms. We assume that firms participating in an auction know the other participants. Each firm must decide whether to cooperate or compete for the contract using two factors: the firm's memory of the previous action of the other firms, and the frequency by which they have met the same firms in the recent past.

For the first factor, the focal firm recalls the previous decision made by the other firms it is meeting using a proportional tit-for-tat strategy~\cite{axelrod1981evolution}. The second factor increases the likelihood of cooperation when the other firms have been. The decision to collude depends on the product of these two: familiarity and experiences of reciprocity are essential to start and sustain collusion~\cite{hilbe2018partners}. In order to keep the model as simple as possible, we do not introduce a price mechanism or consider who wins a given auction. We seek to demonstrate that random spatial correlations can create environments with locations heterogeneously favorable to collusion. We present the precise parameters and initial conditions in the section on data and methods.

\begin{figure}[t]
\includegraphics[width=\textwidth]{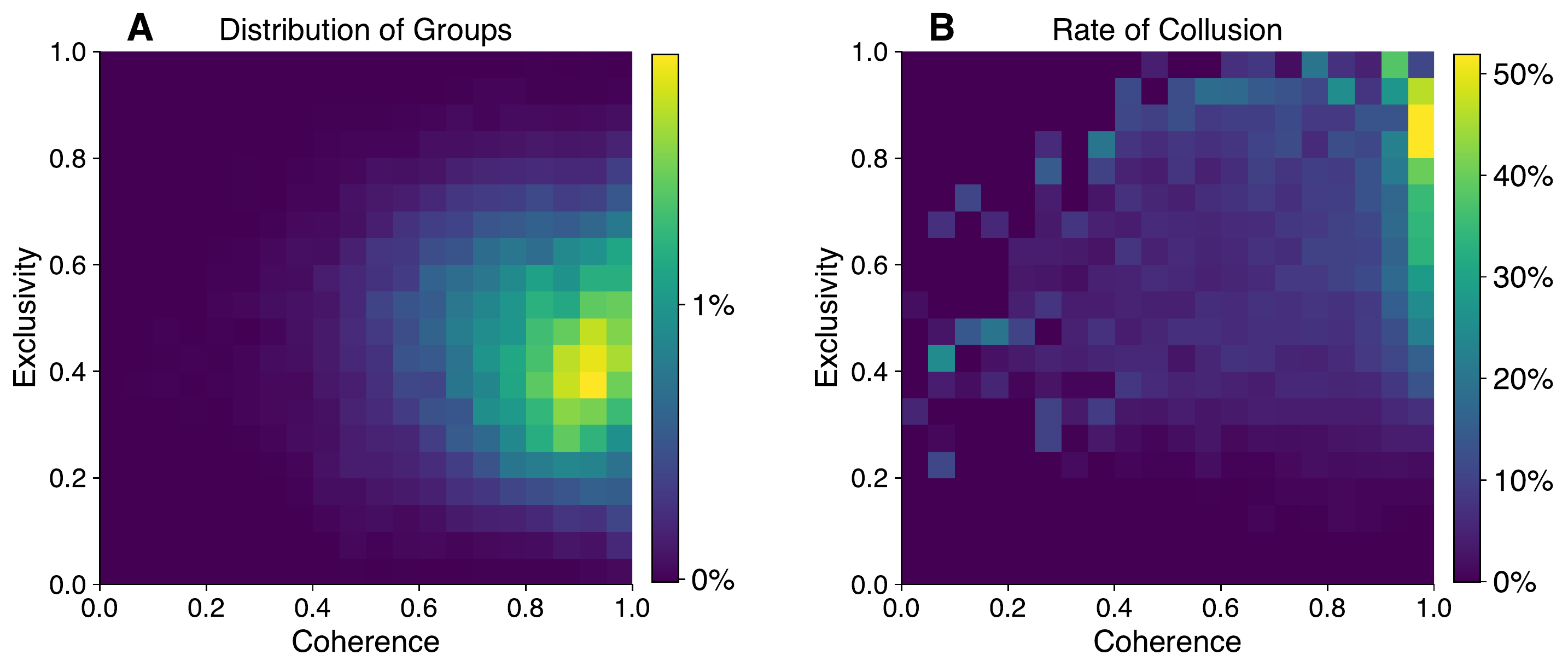}
\caption{Simulation model results of 5000 market simulations. A) the distribution of groups observed from the resulting co-bidding networks in the binned coherence-exclusivity feature space. B) the rate of collusion by groups with given coherence-exclusivity. The model suggests that high coherence and exclusivity groups are not common, but that they have significantly higher rates of collusion.}
\label{fig:simulation_model}
\end{figure}

We simulated 5000 instances of our model, each time initializing a new market with randomly placed firms and contracts. In each instance we award 2,000 contracts, discarding the data from the first 1,000 contracts as burn-in. As before, we constructed the co-bidding networks of firms, detected groups in them, and plotted their distributions in Figure~\ref{fig:simulation_model}, subplot A.

For each group, we calculated the rate at which members unanimously cooperated on a contract, in other words the relatively frequency of successful collusion among the group. We plot the distribution of this frequency across the coherence-exclusivity space in subplot B. In agreement with our empirical evidence, we find that collusion is significantly more likely to emerge among groups in the region of high coherence and exclusivity. These findings are robust to a range of threshold values for agents to cooperate. We report these in the supplementary information.

As discussed earlier in the article, economic theory and empirical observation suggests that there are certain environments in which cartels are more likely to emerge~\cite{levenstein2006determines}. Inspired by the literature on evolutionary game theory, we considered a simple model of cooperation based games played between agents embedded in space~\cite{szamado2008effect}. Our findings support the notion that the co-bidding network captures localized market conditions, which in turn govern the likelihood and effectiveness of emergent cooperation. This indicates that cartels may only be able to survive in the high coherence and high exclusivity zone. Of course this does not imply that any group in this zone will engage in cartel behavior - we present this finding as rather a necessary condition for collusion than a sufficient one. Interestingly, in the case when a market is significantly governed by the locations of firms in physical space, for example in the Ohio milk market, our model has the potential to be calibrated with geographical data.

\section*{Discussion}
In this paper we developed a framework to find groups of firms in public contracting markets and to screen them for collusive markers. Testing our method on a ground truth case, a large scale market without known collusion, and a simple model of such markets, we find that collusion seems more likely to emerge among groups of firms with cohesive and exclusive interactions. Groups occupying such distinguished places in the broader market have found a niche with conditions ripe for the emergence of cooperation.

We must acknowledge that our approach to cartel detection is only suggestive - it cannot prove that a group of firms are engaged in collusion. Our features describe more necessary than sufficient conditions for cartel behavior: patterns of interaction cannot conclusively prove collusion. Rather we propose that our method be used to narrow down a large space of possibilities, into a shorter list of candidates for investigation. Authorities can then apply classical screens for evidence of illegal cooperation~\cite{harrington2006behavioral,fazekas2016}, for example by observing abnormal stability in prices or market shares~\cite{abrantes2006variance,imhof2017simple}, or by comparing observed behavior against a model of competitive behavior~\cite{bajari2003deciding,pesendorfer2000}.  More granular data required for these tests for collusion can be collected once a key subset of firms is identified, at significantly lower cost. Other data-driven studies develop screens for particular exotic auction formats such as average-price auctions or multi-round auctions~\cite{kawai2014detecting} - our approach can also complement these context-specific screens.

Another advantage of our approach is that it does not rely on information from whistleblowers to highlight a candidate group of firms, avoiding a potential source of bias in the cartel literature~\cite{allain2011determination}. We also acknowledge that there are other cartel strategies in public contracting markets beside bid-rigging, for instance when firms agree to stay out of each others' markets entirely.

In our model we do not consider the idea that some firms might simply be honest and refuse to form a cartel even in optimal conditions, nor do we consider how fear of prosecution might influence the choice to collude. Though the illegality of collusion adds an additional obstacle to the emergence of cooperation among firms, the empirical observation that cartel life spans are heterogeneous suggests that many firms are willing to collude, but that only certain environments are conducive to cartels~\cite{levenstein2006determines}. 

In spite of the limitations, we note that our method can be applied to other questions about cartels. For example, what does the co-bidding network look like near a cartel when it is born compared to when it dies? The inner-workings of potential cartels would surely be reflected in network structure of the market. Observed cartels in other contexts have operated by methods such as rotating the winner~\cite{ishii2009favor}, by side payments to losing firms~\cite{pesendorfer2000}, and some even run internal auctions to optimize their profits~\cite{asker2010study}. Observing the relationship between the procedure by which contracts are awarded, for example to the trimmed-average bidder in Italian road contracts~\cite{conley2016} or by randomly chosen open or sealed bid procedures in timber auctions~\cite{athey2011comparing}, and network structure may also reveal whether firms are competing or colluding. The specifics of a market and manner by which contracts are awarded matters a great deal to how collusion might evolve~\cite{hendricks1989}. Certain rules make it easier for firms to collude or easier to detect collusion~\cite{conley2016,athey2011comparing}. 

We are confident that our approach can be applied to these cases in which we have extra information about the rules of a market. It is likely still the case that certain patterns of interaction are effective markers of collusion and that networks provide a useful map of such interactions.

\section*{Methods and Data}

\subsection*{Co-bidding networks, group detection, and group features}
We define a public contracting market's \textit{co-bidding network} as a projection of a bipartite network onto the set of firms active in the market. Specifically, we form a bipartite network of contracts and firms bidding on them, then create a network of firms which bid for the same contract. We weight the connections based on the similarity of co-bidding behavior between firms using Jaccard similarity. Specifically, firm A and firm B are connected by a link with weight equal to the overlap of the contracts they bid on:

$$w_{A,B} = \dfrac{\lvert c_{A} \cap  c_{B}\rvert}{\lvert c_{A} \cup  c_{B}\rvert},$$
where $c_{A}$ ($c_{B}$) is the set of contracts of $A$ ($B$) with at least one other bidder and $|\cdot|$ is the cardinality of a set.

Given a co-bidding network our aim is to extract groups of nodes which may be analyzed for cartel activity. Groups should be communities in the network sense: there should be more interactions within the group than leaving the group. The case in question suggests several other criteria for our algorithm. Groups should be small, as cooperation becomes more difficult to sustain with more participants. Firms might be present in more than one part of the market, so we should consider overlapping groups.

We adapt a bottom-up method for community detection which merges nodes into groups by local optimization of a fitness function from previous work by Lancichinetti, Fortunato, and Kert\'esz (hence: LFK) ~\cite{lancichinetti2009detecting}. We define the fitness $f_{G}$ of a group of nodes $G$ in a network as:

$$f_{G}= \dfrac{s_{in}^{G}}
{\left(s_{in}^{G}+s_{out}^{G}\right)^{\alpha}
\times \lvert G \rvert^{\beta}},$$
where $s_{in}^{G}$ and $s_{out}^{G}$ denote the strength (the sum of weights) of edges within the group and adjacent to the group, respectively. $\lvert G \rvert$ is the size of the group, and $\alpha$ and $\beta$ are free parameters which control the size of the groups found. When $\alpha$ is increased, additional strength is penalized, while $\beta$ penalizes the number of group members independently of their strength. In the paper we set both parameters to 1.5. Increasing $\alpha$ insures that new nodes added to a group interact primarily within the group, while increasing $\beta$ restricts the size of the groups we detect, in line with the stylized facts about cartels from the economics literature that lasting cartels are small and frequently interacting~\cite{levenstein2006determines}. We report the sizes of the groups found in the empirical cases in the SI.

Given such a fitness function of a group of nodes in a co-bidding network, we can define the fitness of a node $n$ relative to a group by calculating the difference in fitness of the group with $n$ and without it:

$$f_{G}^{n}= f_{G+\{ n \}} - f_{G}$$

With this node-level measure of fitness we can define our group detection algorithm. For each node in the network:
\begin{itemize}
    \item select a node $n$ and initialize a group containing only $n$,
    \item select the neighbor of $n$ with the largest fitness and, if it has positive fitness, add it to the group. 
    \item repeat until no nodes adjacent to the group have positive fitness. 
\end{itemize}
In this way we find groups in the network which are overlapping, small (tuned by the parameters), with more weight among themselves than with non-group members. It is possible to save significant computational time by initializing new groups only for nodes that have not been included in a group before. In contrast with the LFK method we do not recalculate the individual fitness of all nodes in the group following the inclusion of a new node. In this sense our adaptation is greedy and not iterative, saving computational time.

Once groups have been extracted from a market's co-bidding network, we then define topological features of each group that may suggest that the firms could form a cartel. The first measure is the \textit{coherence}~\cite{onnela2005intensity} of a group $C_{G}$, the ratio of the geometric and arithmetic means of the edges weights among group members, measuring the balance and overall frequency of interactions among the group members:

$$C_{G}=\dfrac{\displaystyle \left(\prod_{l_{G}}w_{l}\right)^{1/\lvert l_{G}\rvert}}{\dfrac{\displaystyle \sum_{l \in G}w_{l}}{\lvert l_{G}\rvert}}$$

The second measure is \textit{exclusivity} , the ratio of strength within the group over the total strength of the group, excluding on edges to non-group members, measuring the group's relative isolation in the broader market:

$$E_{G}= \frac{s_{in}^{G}}{\left(s_{in}^{G}+s_{out}^{G}\right)}$$

\subsection*{Null models}
In both empirical cases, we created null models of the market to capture the extent to which groups of certain cohesion and exclusivity emerge by chance. For the Ohio school milk data we shuffled the bidders across all contracts, preserving the number of bidders each contract received, and the number of contracts each firm bid on. In Georgia we repeated the same procedure with an additional restriction: firm bids were only shuffled among contracts with the same 2-digit Common Procurement Vocabulary (CPV) code~\cite{cpvreport}. CPV codes describe the type of good or service being contracted, from road repair to medicine. By restricting the random shuffling of bids by CPV code, we create a randomized version of the broader market which preserves the tendency of firms providing similar products to interact.

\subsection*{Agent Based Model}
In this section we describe the specific parameters of our simulated model of a spatially embedded contracting market. Each simulated market was initialized with 50 firms and 75 issuers of contracts (analogous to school districts in the Ohio milk market) placed uniformly at random in the unit square. We then play 2,000 rounds corresponding to contract auctions.

In each round a randomly selected issuer releases a contract $C$ placed nearby (at a position drawn from a 2-d normal distribution centered on the issuer with standard deviation .3). Firms participate in the competition for the contract if they are within .1 distance of the contract (if no firms are close enough, the distance for inclusion is extended by .1 repeatedly until at least one firm participates). The set of firms participating, $F$, is known to all firms.

Each firm must then decide to collude or compete. Collusion is successful if all firms collude. Each firm $f$ considers two pieces of information about the other firms in its decision making process, its \textit{memory} of previous interactions with each other firm, and the relative \textit{frequency} with which it meets with the others. It recalls the decision made by the other firms the previous time they met (initialized randomly) and calculates the share of previous round cooperators:

$$f_{memory}=\dfrac{ \sum_{\hat{f} \in F\setminus f} 
  \delta_{\hat{f}}^{C_{prev}}}
{\|F\setminus f \|}, $$

where $\delta_{\hat{f}}^{C_{prev}}$ equal 1 if $\hat{f}$ cooperated the last time it encountered $f$. This is the proportional (compared to the absolute) generalization of the tit-for-tat strategy to multi-agent games~\cite{axelrod1981evolution}.

Next, $f$ considers how often, in the last $k$ contracts it was participating in, the current other firms were a subset of the participating firms. If this is true at least two-thirds of the time, the firm considers the other firms it meets as familiar.

$$ f_{frequency} = 
    \begin{cases}
      1, & \text{if}\ \dfrac{\sum_{i=1}^{k} \delta_{F \subset F(C_{f}^{i})}}{k} \geq 2/3 \\
      0, & \text{otherwise}
    \end{cases}
$$

where $F(C_{f}^{i})$ denotes the firms participating in the $i$'th previous contract of firm $f$. $f_{frequency}$ increases as $f$ tends to meet the same firms. We set $k$ to 10 in the paper. 

The focal firm's decision to collude or compete depends on the product of these two factors:

$$f_{compete} =  
    \begin{cases}
      1, & \text{if}\ f_{memory} * f_{frequency} > .9 \\
      0, & \text{otherwise}
    \end{cases}
$$

Finally, we add noise to the system by allowing a .1\% chance that a firm spontaneously colludes. In our model agents do not learn or track the outcome of their actions - they only react to their most recent memory of other firms and the frequency by which they meet. After 2,000 contracts are awarded, we end the simulation and discard the outcomes of the first 1,000 contracts as burn-in.

\subsection*{Datasets}
The Ohio school milk data was generously provided by Porter and Zona~\cite{porter1999ohio}. The data consists of a significant share of all school-milk procurement contracts from 1980s Ohio provided to Porter and Zona by the State of Ohio. Porter and Zona served as expert witnesses in a trial against the suspect cartel. There are several other significant examples of cartels in public school milk markets in the US during the 1980s, for example in Florida and Texas~\cite{pesendorfer2000,hewitt1993incumbency}.

We collected the Georgian contracts dataset from the centralized procurement portal of the State Procurement Agency (SPA) of Georgia (https://tenders.procurement.gov.ge/), including all contracts awarded through the portal between 2011 and 2016. Contracts are awarded to the lowest bidder in a sealed-bid auction. Each contract includes a product category (CPV code~\cite{cpvreport}, which we use for the null model, and a reserve price, the maximum price that the public buyer would pay for the good or service, which we use to normalize prices. The procurement portal also reports bid protests: these are legal disputes of participants in the procurement process against the agency issuing a contract. For example, a firm may protest that it was unfairly excluded from the competition for a contract.

\subsection*{Acknowledgements}
The authors thank Robert Porter for sharing the Ohio school milk dataset. We also acknowledge Greg Taylor, Roberta Sinatra, and Federico Battiston for helpful comments. JW acknowledges the support of a Swedish Competition Authority grant (KKV 72/2018-Ab-1). JK acknowledges support from the Hungarian Scientific Research Fund (OTKA K129124 - "Uncovering patterns of social inequalities and imbalances in large-scale networks").

\subsection*{Author contributions statement}
JW and JK conceived of the idea of the paper and developed the methods. JW collected the data and implemented the methods. JW and JK and wrote the manuscript.

\subsection*{Competing interests}
The authors declare no competing interests.

\subsection*{Corresponding author}
Correspondence to johanneswachs@gmail.com.

\subsection*{Data Availability Statement}
Data to reproduce our analysis available upon request.

\bibliographystyle{acm}
\bibliography{cartels.bib}

\newpage

\section*{Supplementary Information}

\section*{Ohio School Milk Data}
\begin{figure}
\centering
\includegraphics[width=\textwidth]{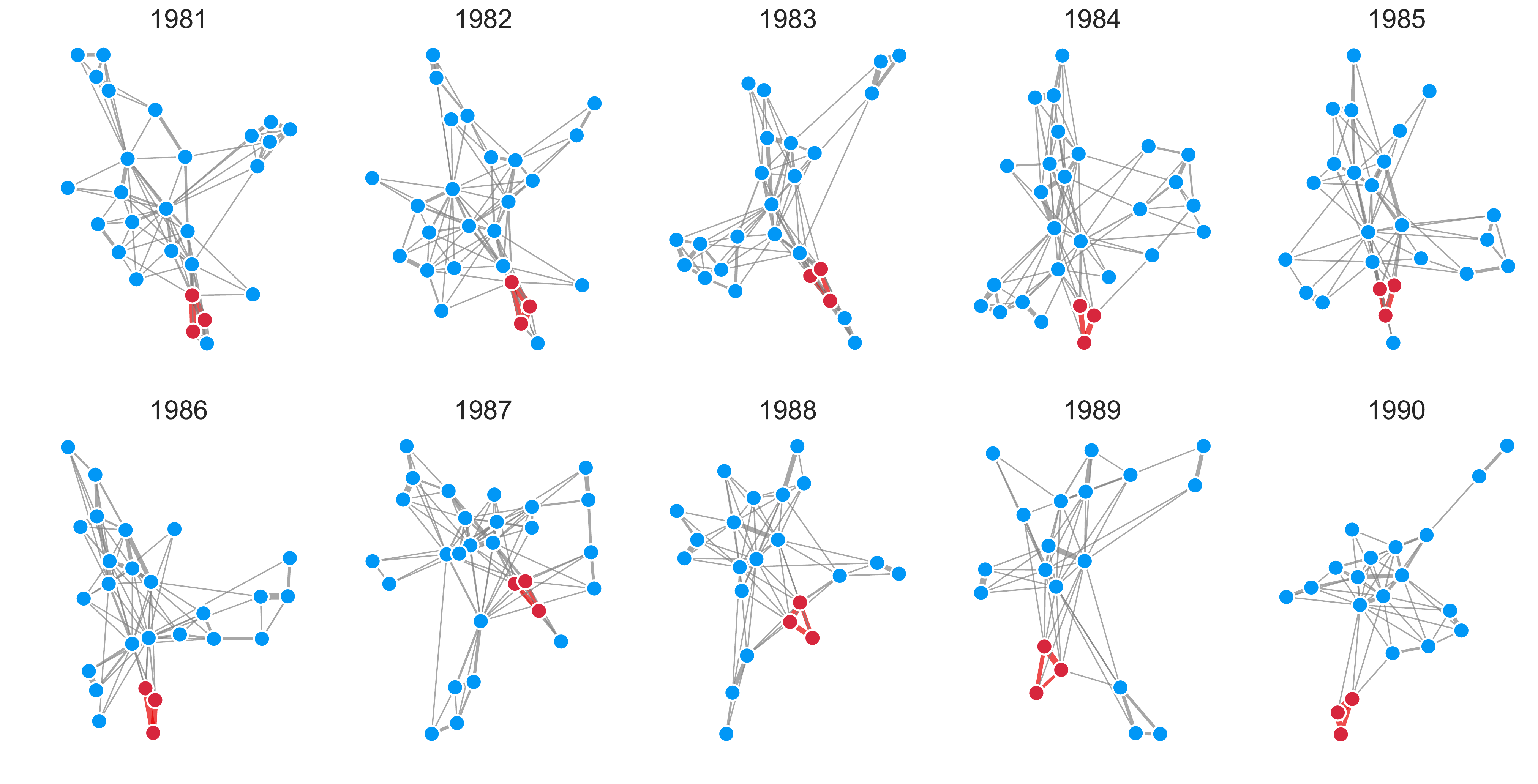}
\caption{Ohio school milk procurement market co-bidding networks, 1981-1990. Red nodes are members of the alleged cartel. For the purposes of visualization we filter out nodes participating in fewer than 5 auctions with other firms. Nodes are placed using a force-layout algorithm, with initial position equal to the final position of the nodes in the previous year.}
\label{fig:ohio_nets}
\end{figure}

\begin{figure}
\includegraphics[width=\textwidth]{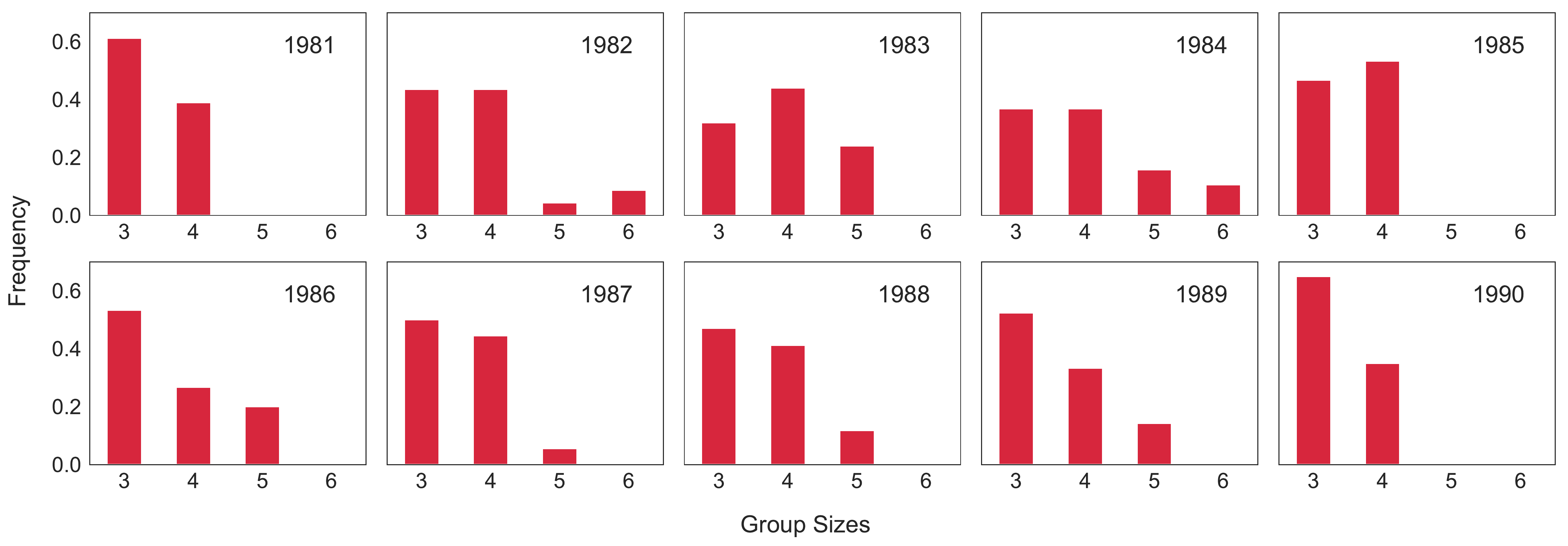}
\caption{Distributions of detected group sizes from the Ohio school milk contracting data, by year.}
\label{fig:oh_group_sizes}
\end{figure}

In this section of the appendix we report summary statistics about the Ohio school milk market in Table~\ref{tab:app_ohio_summary} and plot the co-bidding networks of the market annually in Figure~\ref{fig:ohio_nets}. We plot the distributions of the sizes of groups detected in the co-bidding network by year in Figure~\ref{fig:oh_group_sizes}.

\begin{figure}
\centering
\includegraphics[width=\textwidth]{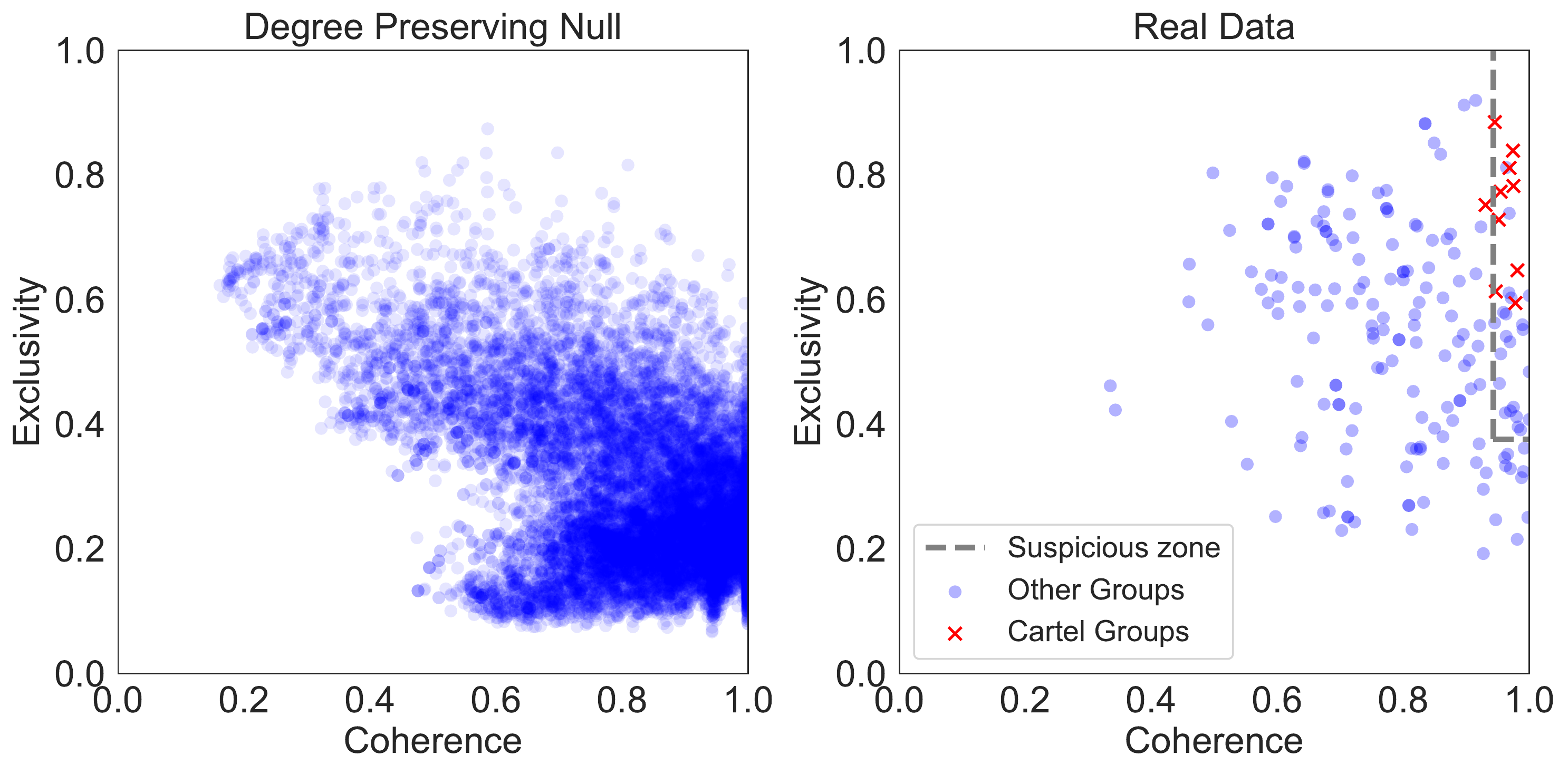}
\caption{Groups detected from the co-bidding network representation of the Ohio school milk procurement market, 1981-1990, placed by their coherence and exclusivity. The data on the left is generated from 100 degree-preserving randomizations per year. On the right we display the groups extracted from the empirical data, with the cartel group in red. We define a suspicious zone, as done for the Georgian data in the primary paper.}
\label{fig:ohio_scatter}
\end{figure}

We plot the observed and null model groups embedded in the coherence-exclusivity space as a simple scatter plot in Figure~\ref{fig:ohio_scatter}, to complement the 2-D histogram reported in the main paper. We include the suspicious zone, analogous to our approach to the Georgian data,  for this dataset, defined as the minimum of the 80th percentile coherence and exclusivity values observed across the 10 years of the dataset.

We use this suspicious zone to calculate the accuracy of our method, including the share of false-positives: non-cartel groups our methods deems suspicious. Out of 186 groups detected in the 10 year period, 31 are labeled suspicious. 9 of the 10 cartel instances are correctly labeled as suspicious for a true positive rate (\textit{recall}) of 90\%. 22 of 176 non-cartel groups are labeled as suspicious, for a false positive rate of 12.5\%. This latter error rate may be overestimated, however, as we do not know if these groups were competing or colluding. 

\begin{table}[b]
\ra{1}
\begin{tabular}{lccc}
\toprule
\setlength{\tabcolsep}{3em}
Year & \# Contracts & \# Firms & Avg. \# Bidders   \\ 
\midrule
1981 & 273   & 49 & 1.96   \\
1982 & 287  & 43 & 1.96 \\
1983 & 318   & 46 & 2.01   \\
1984 & 339  & 55 & 1.94 \\
1985 & 357   & 49 & 1.93   \\
1986 & 378  & 49 & 2.02 \\
1987 & 411   & 42 & 1.97   \\
1988 & 419  & 41 & 1.83 \\
1989 & 392   & 40 & 1.76   \\
1990 & 331  & 43 & 1.73 \\
\bottomrule
\end{tabular}
\caption{Summary statistics of the Ohio school milk market by year.}
\label{tab:app_ohio_summary}
\end{table}

\section*{Georgian Contracting Data}
\begin{figure}
\includegraphics[width=\textwidth]{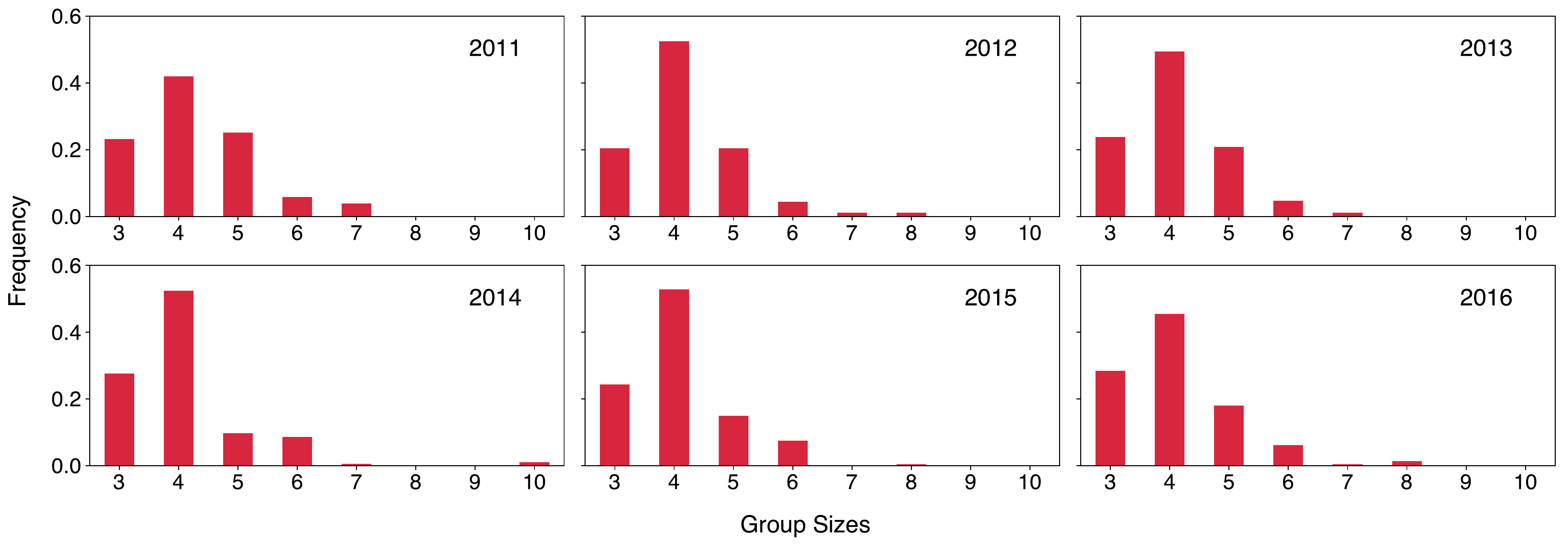}
\caption{Distributions of detected group sizes from the Georgian contracting data, by year.}
\label{fig:ga_group_sizes}
\end{figure}

In Table~\ref{tab:app_georgia_summary} we report annual summary statistics on the Georgian public procurement market. In Figure~\ref{fig:ga_group_sizes} we plot the distribution of the sizes of the groups we detected in the Georgian co-bidding network each year.

\begin{table}[b]
\ra{1.2}
\begin{tabular}{lcccccccc}
\toprule
Year & \# Contracts & \# Firms & Avg. \# Bidders & Share Protested & Billion GEL  & Avg. Rel. Cost    \\ \midrule
2011 & 17,396   & 3,804 & 1.73 & .003& 1.33 & .88 & \\
2012 & 18,575  & 4,048 & 1.75 & .004&1.33& .88 & \\
2013 & 20,230   & 4,399 & 2.03& .012 & 1.67 & .85 &\\
2014 & 22,122  & 4,884 & 2.02& .014 &1.81 & .86 & \\
2015 & 26,033   & 5,600 & 2.02& .023 & 2.30 & .87 & \\
2016 & 28,092  & 6,191 & 2.15& .031 &2.50 & .87 & \\
\bottomrule
\end{tabular}
\caption{Summary statistics of the public contracting market of the Republic of Georgia by year. Share protested refers to the share of contracts legally protested by firms, Avg. Rel. Cost and StDev. Rel. Cost refer to the average cost of a contract, scaled by the maximum reserve price, and the standard deviation of the same, respectively. 1 Georgian Lari (GEL) equals roughly .6 US Dollars from 2011-2014, then  .45 in 2015-2016.}
\label{tab:app_georgia_summary}
\end{table}

\begin{figure}
\includegraphics[width=0.65\textwidth]{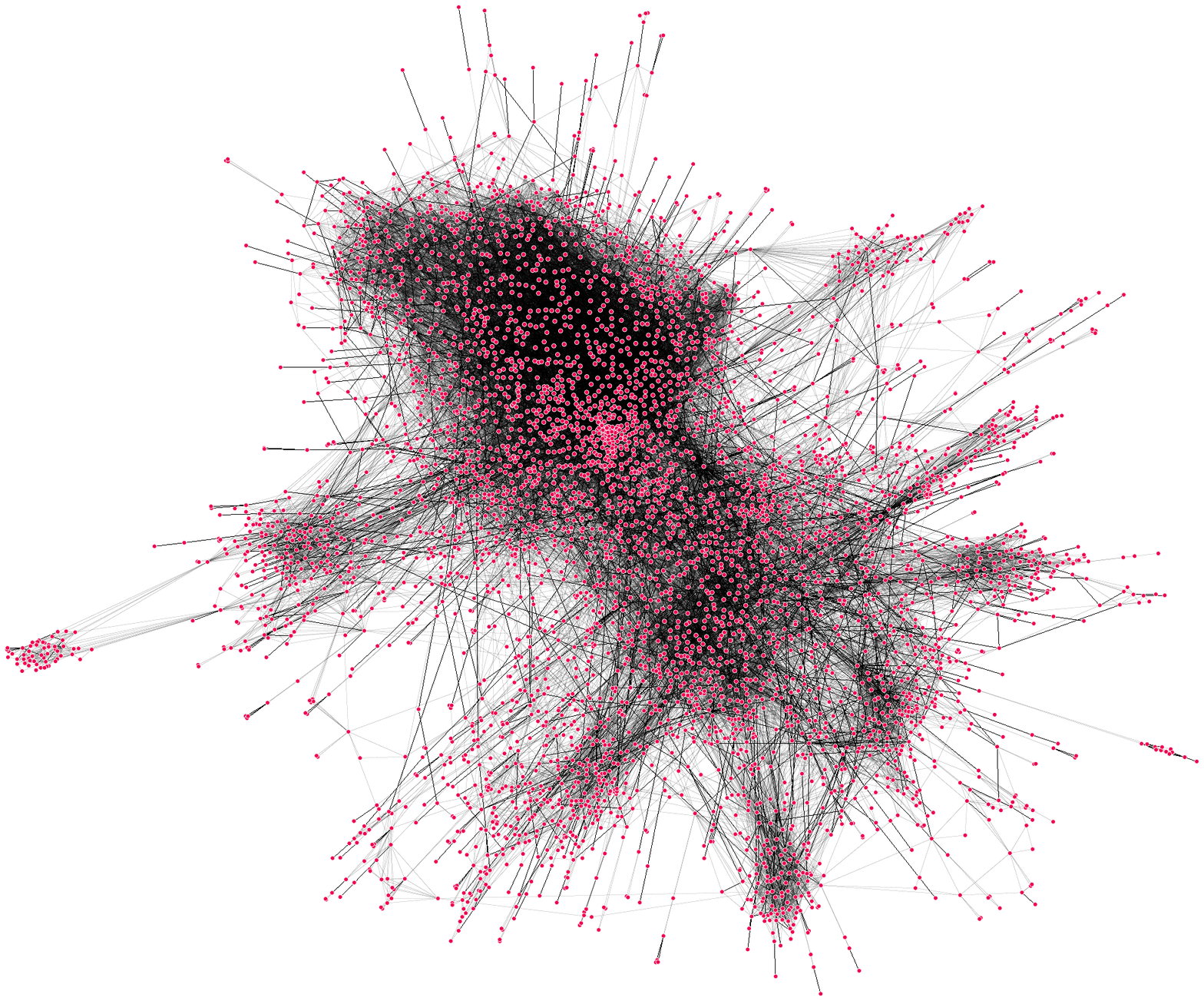}
\caption{The giant connected component of the 2013 Georgian public procurement market's co-bidding network, containing nearly 5,000 firms which won over 90\% of all contracts awarded that year. Nodes, representing firms, are connected by an edge if they bid on the same contract in 2013.}
\label{fig:ga_2013_net}
\end{figure}

We visualize the full co-bidding network of 2013 in Figure~\ref{fig:ga_2013_net}. Though there are some distinct clusters of firms, this visualization of the dense connections between firms suggests that segmenting the market by product would ignore the fact that firms seem to participate in diverse markets with otherwise distant firms.

\begin{table}[]
\ra{1}
\begin{tabular}{lrrcrrclr}
& \multicolumn{2}{c}{Suspicious Groups} & \phantom{abc}& \multicolumn{2}{c}{Ordinary Groups} & \phantom{abc} & \multicolumn{2}{c}{Differences} \\
\cmidrule{2-3} \cmidrule{5-6}  \cmidrule{8-9} 
\multicolumn{1}{l}{\textbf{Threshold = 20}}  &\textit{Mean} & \textit{Std. Dev.} &&\textit{Mean} & \textit{Std. Dev.} &&\textit{M-W U} & \textit{p-value}\\ \midrule

 Avg. Relative Price & 0.938   & 0.046 && 0.914 & 0.053 &&$30211^{***}$ & $<$0.001 \\
 Avg. $CV_{price}^{G}$ & 0.098  & 0.055 && 0.117 & 0.059 && $33470^{***}$ & $<$0.001 \\
 Avg. $CV_{bidding}^{G}$ & 0.047   & 0.056 && 0.055 & 0.038 && $32306^{***}$ & $<$0.001  \\
 In-group Bid Protest Rate & 0.134   & 0.341 && 0.237 & 0.425 && $37516^{*}$ & 0.011 \\

\end{tabular}

\begin{tabular}{@{}lrrcrrclr@{}}
& \multicolumn{2}{c}{\phantom{abc}} & \phantom{abc}& \multicolumn{2}{c}{\phantom{abc}} & \phantom{abc} & \multicolumn{2}{c}{\phantom{abc}} \\
 \multicolumn{1}{l}{\textbf{Threshold = 10}}&\textit{Mean} & \textit{Std. Dev.} &&\textit{Mean} & \textit{Std. Dev.} &&\textit{M-W U} & \textit{p-value}\\ \midrule
 Avg. Relative Price & 0.955   & 0.045 && 0.924 & 0.056 && $140930^{***}$ & $<$0.001  \\
 Avg. $CV_{price}^{G}$ & 0.075  & 0.067 && 0.105 & 0.068 && $158212^{***}$ & $<$0.001 \\
 Avg. $CV_{bidding}^{G}$ & 0.034   & 0.062 & & 0.050 & 0.056 & & $167427^{***}$ & $<$0.001 \\
 In-group  Bid Protest Rate & 0.077   & 0.266 & & 0.113 & 0.317 & & 226584 & 0.0791  \\
\end{tabular}

\begin{tabular}{@{}lrrcrrclr@{}}
& \multicolumn{2}{c}{\phantom{abc}} & \phantom{abc}& \multicolumn{2}{c}{\phantom{abc}} & \phantom{abc} & \multicolumn{2}{c}{\phantom{abc}} \\
 \multicolumn{1}{l}{\textbf{Threshold = 5}}&\textit{Mean} & \textit{Std. Dev.} &&\textit{Mean} & \textit{Std. Dev.} &&\textit{M-W U} & \textit{p-value}\\ \midrule
 Avg. Relative Price & 0.975   & 0.039 && 0.925 & 0.056 && $49270^{***}$ & $<$0.001  \\
 Avg. $CV_{price}^{G}$ & 0.066  & 0.054 && 0.104 & 0.068 && $49663^{***}$ & $<$0.001 \\
 Avg. $CV_{bidding}^{G}$ & 0.023   & 0.038 & & 0.050 & 0.057 & & $50594^{***}$ & $<$0.001 \\
 In-group  Bid Protest Rate & 0.132   & 0.339 & & 0.111 & 0.314 & & 80909 & 0.0791  \\
\end{tabular}
 \caption{Robustness check of cartel screens applied to suspicious and ordinary groups of firms detected in the Georgia procurement market, 2011-2016. We vary the threshold of the minimum number of contracts bid on exclusively by members of the group (20, 10, 5, compared with 30 in the main text). We replicate the main findings in the text that cartel groups have higher average relative prices, and are more likely to have a low average coefficient of variation on bids for a contract. The finding that suspicious groups are more likely to legally protest the winnings of other group members is no longer statistically significant when we filter at 5 or 10 contracts. Mann-Whitney U: * $p < .05$, ** $p <.01$,*** $p <.001$ }
\label{SI:ga_stats_table_threshold}
\end{table}

\section*{Simulation model}
\begin{figure}
\includegraphics[width=\textwidth]{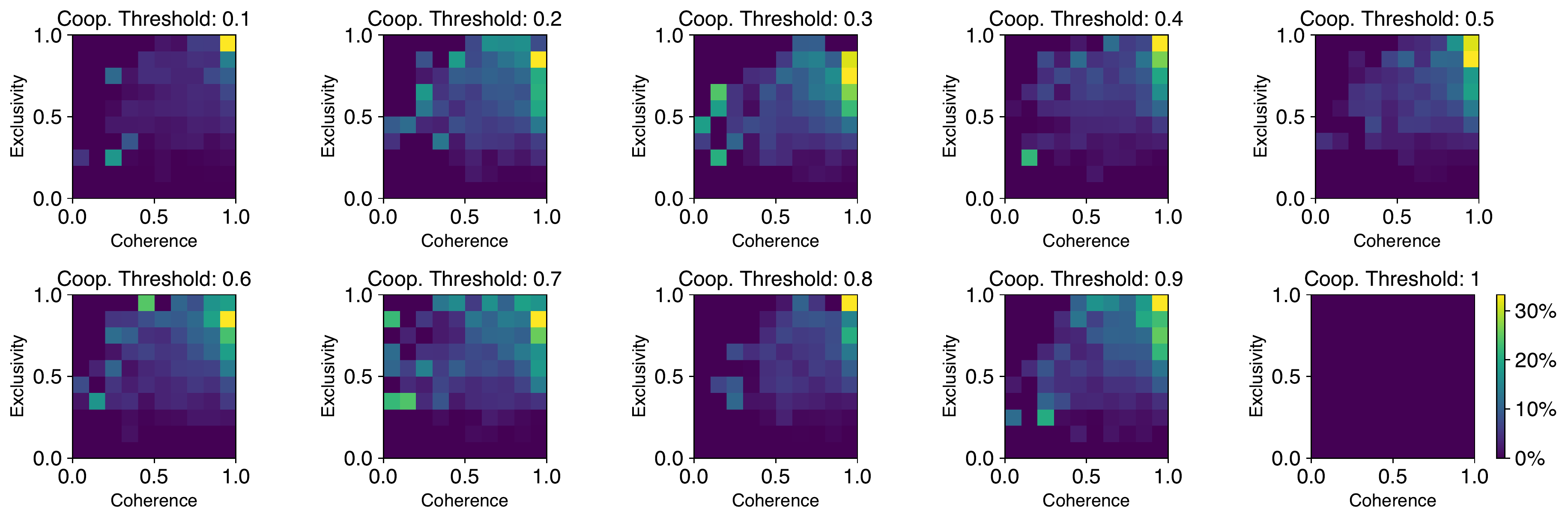}
\caption{The emergence of collusion in the simulation model as the threshold for agent cooperation varies. We find that for a wide range of the threshold parameter groups in the high-coherence, high-exclusivity are significantly more like to successfully cooperate.}
\label{fig:sim_robust}
\end{figure}

In order to assess the sensitivity of our simulation model to the choice of parameters, we carried out several robustness tests. The key parameter, $f_{compete}$, the threshold at which an agent decides to cooperate given the frequency of past encounters and past cooperation which is set to .9 in the paper, is varied between 0 and 1, increasing in steps of .1. We repeat the simulation described in the main paper for  and plot the 2-d histogram of cooperation rates by coherence-exclusivity bin in Figure~\ref{fig:sim_robust}. We find that for all threshold cooperation is more likely to emerge in the high-coherence and high-exclusivity regime, as expected.

\end{document}